\documentclass[prd,twocolumn,floatfix,amsmath,nofootinbib,amssymb,floatfix]{revtex4}
\usepackage{graphicx,color,dcolumn,booktabs,bm}
\usepackage{longtable,lscape}
\usepackage{pdfpages}
\usepackage{txfonts}
\usepackage{overpic}
\usepackage{amssymb}
\usepackage{makecell}
\usepackage{indentfirst}
\usepackage{feynmf}   %{feynmp}
\usepackage{slashed}  %for Feynman symbols
\usepackage{cases}
\usepackage{color}
\usepackage{multirow}
\usepackage{threeparttable}
\usepackage{epstopdf}
\usepackage{enumerate}
\usepackage{subfigure}
\usepackage{diagbox}
\usepackage{graphicx,color,dcolumn,booktabs,bm}
\usepackage{mathrsfs}
\usepackage{cancel}
\usepackage{float}
\usepackage[colorlinks,
            citecolor=blue,
            anchorcolor=red,
            menucolor=red,
            linkcolor=red,
            filecolor=red,
            runcolor=red,
            urlcolor=blue,
            frenchlinks=red]{hyperref}

\begin{document}
\title{Heavy quark dominance in orbital excitation of singly and doubly heavy baryons }
\author{Zhen-Yu Li$^{1}$}
\email{zhenyvli@163.com }
\author{Guo-Liang Yu$^{2}$}
\email{yuguoliang2011@163.com }
\author{Zhi-Gang Wang$^{2}$ }
\email{zgwang@aliyun.com }
\author{Jian-Zhong Gu$^{3}$ }
\email{jzgu1963@ciae.ac.cn }

\affiliation{$^1$ School of Physics and Electronic Science, Guizhou Education University, Guiyang 550018,
China\\$^2$ Department of Mathematics and Physics, North China Electric Power University, Baoding 071003,
China\\$^3$ China Institute of Atomic Energy, Beijing 102413,China}
\date{\today }

\begin{abstract}
A mechanism of the heavy quark dominance in the orbital excitation is proposed in this paper which is testified to be reasonable for singly and doubly heavy baryons.
In the relativistic quark model, an analysis of the Hamiltonian figures out the mechanism that the excitation mode with lower energy levels is always associated with the heavy quark(s), and the splitting of the energy levels is  suppressed by the heavy quark(s). So, the heavy quarks dominate the orbital excitation of singly and doubly heavy baryons. Furthermore, a physical understanding of this mechanism is given in a semi-classical way.
Accordingly, the predicted mass spectra of singly and doubly heavy baryons confirm the rationality of this mechanism. In addition, an interesting consequence of this mechanism is that a heavy-light meson is more likely to be produced in the strong decay of the high-orbital excited states, which is supported by experiments. This mechanism is rooted in the breakdown of the mass symmetry. Therefore, it may be also valid for other multi-quark systems, such as the tetraquarks $Qqqq$ and $QQqq$, or the pentaquarks $Qqqqq$ and $QQqqq$.

\end{abstract}


\maketitle

\section{Introduction}\label{sec1}

Quantum Chromodynamics (QCD), the theory of quarks
and gluons, has been developed for 50 years\cite{f01}. An important prediction of QCD is the existence of heavy baryons. The study of heavy baryon spectroscopy has always been one of the hot topics. It contributes to a deeper understanding of the properties of heavy quarks, such as the heavy quark symmetry, chiral dynamics, feature of strong interactions, and relevant models inspired by QCD.
Until now, there have been various methods to analyze baryon spectra (see~\cite{f01,fp01} and Refs. therein).
Among them, quark potential models are currently an indispensable tool for guiding our complete understanding of the multitude of observed baryons and their properties\cite{f02}.

Copley, Isgur and Karl first studied the charmed baryons in a quark potential model with hyperfine interactions\cite{f03}. While the first prediction of the full spectrum of baryons including heavy-quark baryons was presented by Capstick and Isgur in 1986, where the relativistic quark model was developed\cite{f04,f05}. Twenty years later, Roberts and Pervin analyzed the heavy baryons systematically and exhaustively, using a non-relativistic quark model\cite{f06}.
All of the above studies tried to give a unified theoretical description of all baryons including heavy-quark ($Q$) baryons, by solving the three-quark system explicitly. These works continue to serve as guidelines for experimental and theoretical researches.
However, their calculations predict more heavy-quark baryon excited states than those presented in the spectra extracted from data. It is actually a general problem with baryon spectroscopy, which first appeared in the light-quark ($q$) baryons and was called the `missing resonance' problem\cite{f01,fp01,f02}.

A possible solution to this problem is to partially freeze the dynamical degrees of freedom in a three-quark system. For example, if a baryon consists of a quark and
a diquark, the reduction of the number of internal degrees of freedom would lead to a more sparsely populated spectrum\cite{f02}. Based on this idea, Ebert, Faustov and Galkin calculated the spectra of the singly heavy baryons in the heavy quark-light diquark picture in the framework of the QCD-motivated relativistic quark model, where the excitations between the two light quarks are frozen\cite{f07,f07p}. It was found that the available experimental data can be well described and the model predicts significantly fewer states than the genuine three-body picture\cite{f07p}. Later, some theoretical works were carried out to investigate the singly heavy baryons with similar methods\cite{f08,f08p,f08p2,f08p3,f08p4}. These methods have also been applied to the spectroscopy of doubly heavy baryons, where the light quark-heavy diquark structure is assumed\cite{f09,f09p1,f09p2,f09p3,f09p4}. However, the reliability of the extension of this method has yet to be tested. `It is telling that this simple diagnostic is difficult to apply since so little is known of the excited baryon spectrum.'\cite{f01}

\begin{figure}[htbp]
\centering
\includegraphics[width=8.5cm]{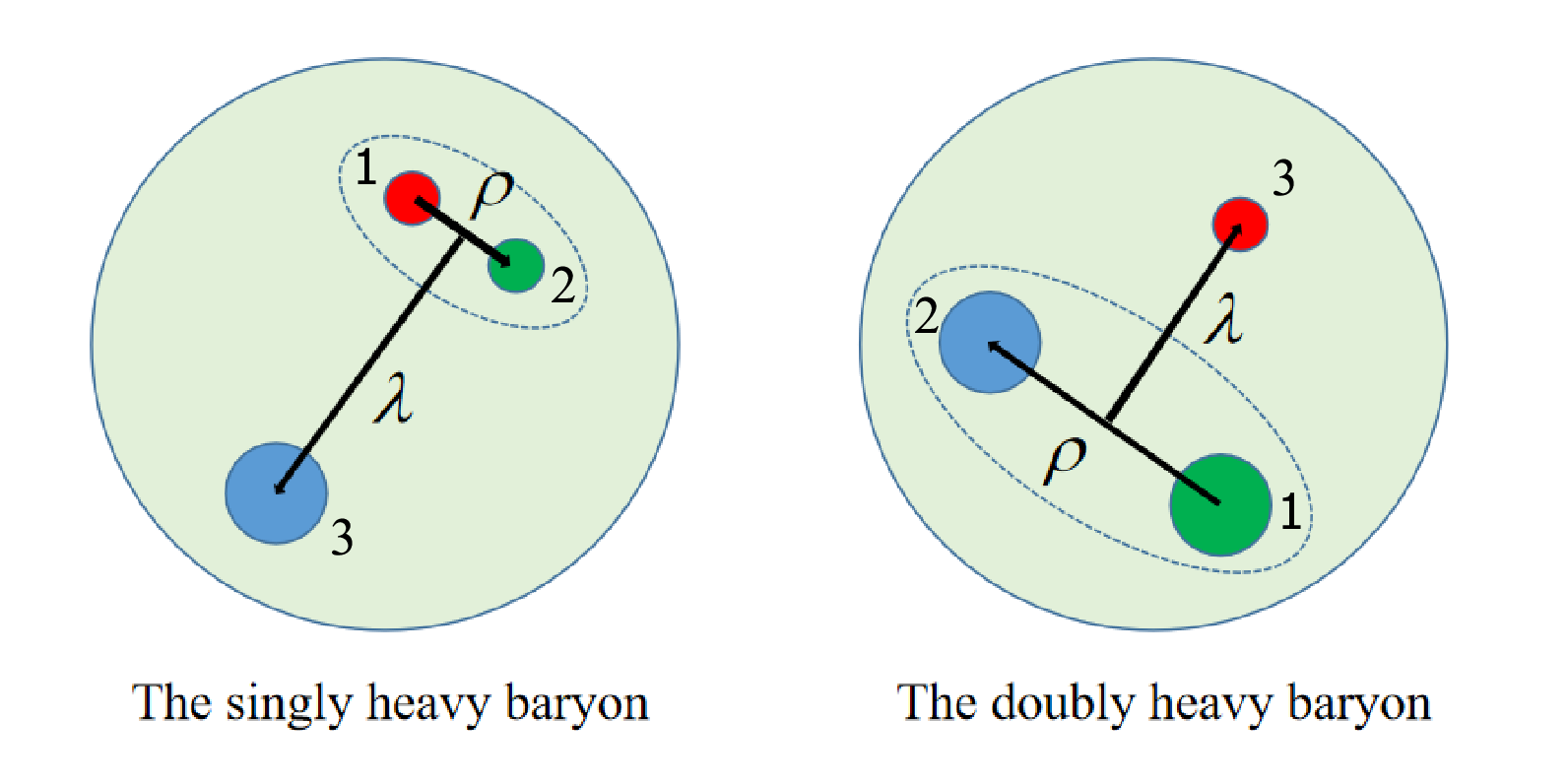}
\caption{Jacobi coordinates for the three-quark systems. The heavy quarks are represented by bigger balls and the light quarks by smaller balls. They are numbered for ease of use in calculations.}
\label{f1}
\end{figure}

For a three-quark system, quark models feature the orbital excitations in two relative coordinates. As shown in Fig.\ref{f1}, they are often taken
to be the Jacobi coordinates $\boldsymbol{\rho}$ and $\boldsymbol{\lambda}$, which is consistent with the flavor sectors of the light quark pair ($q_{1}q_{2}$) in a singly heavy baryon or the heavy quark pair ($Q_{1}Q_{2}$) in a doubly heavy baryon (see Sect.\ref{sec2} A).
In 2016, Yoshida, Hiyama, Hosaka, Oka and Sadato dealt rigorously with the three-quark systems, and analyzed the low-lying excitations of singly and doubly heavy baryons, respectively\cite{f10}.
They concluded that the excitation of the $\lambda$-mode becomes dominant in low-lying states of singly heavy baryons. While the
 $\rho$-mode becomes dominant in low-lying states of doubly heavy baryons. In fact, these features have been mentioned earlier\cite{f03,f06}, and in particular, they were expressed systematically in Ref.\cite{f06}. Later, Chen, Luo, Liu and Matsuki noticed that almost all observed single heavy baryons could be explained as the $\lambda$-mode excited states (in the diquark picture) in theory\cite{f10p}.

Inspired by the above theoretical achievements, in 2022, the spectra of singly heavy baryons were studied systematically, in which all orbital excitations were assumed in the $\lambda$-mode\cite{f11,f11p} and the $\rho$-mode was used alone to study the excitations of doubly heavy baryons\cite{f12,f12p,f12p2} in the framework of the relativistic quark model\cite{f04,f05}. The calculated results are satisfactory, especially, the predicted spectra of singly heavy baryons can reproduce almost all experimental data nicely. Then, the rationality of this approach and the physical nature behind it need to be carefully considered.

 As shown in Fig.\ref{f1}, the determination of the Jacobi coordinates $\boldsymbol{\lambda}$ in a singly heavy baryon and $\boldsymbol{\rho}$ in a doubly heavy baryon is associated with the heavy quark(s). The same is true of the corresponding orbital excitation modes. Hence, it implies that the orbital excitation may be dominated by the heavy quarks.
In this paper, we will comprehensively analyze the heavy quark dominance (HQD) mechanism in orbital excitations of singly and doubly heavy baryons, and discuss the reasonability of the HQD approximation.

The remainder of this paper is organized as follows. In
Sect.\ref{sec2}, the HQD mechanism is analyzed with the Hamiltonian of the relativistic quark model and a semi-classical approach. In Sect.\ref{sec3}, the HQD approximation is tested in singly heavy baryons. Sect.\ref{sec4} discusses the HQD approximation in doubly heavy baryons. The signature of the HQD mechanism in the strong decay is analyzed in Sect.\ref{sec5}. And Sect.\ref{sec6} is reserved for the conclusions.

\section{Analysis of the HQD approximation with the Hamiltonian}\label{sec2}

\subsection{Hamiltonian and the wave function}\label{sec2.1}
In our previous studies on the spectroscopy of singly and doubly heavy baryons, we adopted the relativistic quark model~\cite{f04,f05}. In the calculation, the Gaussian  expansion  method  (GEM)  and  the  infinitesimally-shifted Gaussian (ISG) basis functions have been used as well~\cite{f12p3}. For the convenience of discussion, a brief introduction of the Hamiltonian and the wave function is presented here.
In the relativistic quark model, the Hamiltonian for a three-quark system reads,
\begin{eqnarray}
\notag
H &&=H_{0}+H^{conf}+H^{so}+H^{hyp}\\
&&=\sum_{i=1}^{3}\sqrt{p_{i}^{2}+m_{i}^{2}}+\sum _{i<j}(H^{conf}_{ij}+H^{so}_{ij}+H^{hyp}_{ij}),
\end{eqnarray}
where the interaction terms $H^{conf}_{ij}$, $H^{so}_{ij}$ and $H^{hyp}_{ij}$ are the confinement, spin-orbit and hyperfine interactions, respectively. The confinement term $H^{conf}_{ij}$ includes the one-gluon exchange potentials $G_{ij}$ and linear confining potentials $S_{ij}$, in which
the smearing transformations are performed. And the interactions have been modified with the momentum-dependent factors.

With the requirement of the flavor $SU(3)_{F}$ subgroups for the light quark pair ($q_{1}q_{2}$), the singly heavy baryons belong to either a sextet $(\mathbf{6}_{F})$ of the flavor symmetric states, or an antitriplet $(\bar{\mathbf{3}}_{F})$ of the flavor antisymmetric states.  Thus, $\Lambda_{Q}$ and $\Xi_{Q}$ belong to the $\bar{\mathbf{3}}_{F}$ sector; $\Sigma_{Q}$, $\Xi_{Q}^{'}$ and $\Omega_{Q}$ belong to the $\mathbf{6}_{F}$ sector. Here $Q$ denotes charm ($c$) quark or bottom ($b$) quark. Analogously, in the doubly heavy baryons, $\Xi_{bc}$ and $\Omega_{bc}$ belong to the $\bar{\mathbf{3}}_{F}$ sector, $\Xi_{bb(cc)}$, $\Xi_{bc}^{'}$, $\Omega_{bb(cc)}$ and $\Omega_{bc}^{'}$ belong to the $\mathbf{6}_{F}$ sector, with the requirement of the heavy quark pair ($Q_{1}Q_{2}$) as shown in Fig.\ref{f1}.

Based on the above discussion and the heavy quark effective theory (HQET)~\cite{f06,f12p4,f12p5}, the spin and orbital wave function of the state is written as
\begin{eqnarray}
\notag
|l_{\rho} \ l_{\lambda} \ L \ s\ j\ J\ M_{J}\rangle &&= \{[(|l_{\rho}\ m_{\rho} \rangle |l_{\lambda}\ m_{\lambda} \rangle)_{L}\times(|s_{1}\ m_{s_{1}} \rangle|s_{2}\ m_{s_{2}} \rangle)_{s}]_{j}\\
&&\times|s_{3}\ m_{s_{3}} \rangle \}_{J M_{J}}.
\end{eqnarray}
$l_{\rho}$($l_{\lambda}$), $L$ and $s$ are the quantum numbers of the relative orbital angular momentum $\textbf{\emph{l}}_{\rho}$ ($\textbf{\emph{l}}_{\lambda}$), total orbital angular momentum $\textbf{\emph{L}}$, and total spin of the quark pair $\mathbf{s}$, respectively. $j$ denotes the quantum number of the coupled angular momentum of $\textbf{\emph{L}}$ and $\textbf{s}$, so that the total angular momentum $J=j\pm\frac{1}{2}$. And $M_{J}$ is the 3rd component of $\textbf{\emph{J}}$. Then, the baryon state is simply labeled with $nL(J^{P})_{j}$, in which $n$ is the quantum number of the radial excitation.

The orbital part of the above wave function $(|l_{\rho}\ m_{\rho} \rangle |l_{\lambda}\ m_{\lambda} \rangle)_{L}$ is expanded in terms of the Gaussian basis functions. Then, the matrix element of each Hamiltonian term can be obtained on these bases with the Jacobi coordinates $\boldsymbol{\rho}$ and $\boldsymbol{\lambda}$ as shown in Fig.\ref{f1} (for details, see~\cite{f04,f11,f11p}). Thus, the eigenvalues of the Hamiltonian for the orbital excited states can be calculated. Particularly, the contribution of each term to the eigenvalues can be analyzed, which will be done in the next subsection.

\subsection{HQD mechanism in the orbital excitation}\label{sec2.2}

For the $L$-wave excitation with $\textbf{\emph{L}}$=$\textbf{\emph{l}}_{\rho}$+$\textbf{\emph{l}}_{\lambda}$, there are an infinite number of orbital excitation modes.
Taking $L=1$ as an example, the excitation modes $(l_{\rho},l_{\lambda})_{L}$ are $(1,0)_{1}$, $(0,1)_{1}$, $(1,1)_{1}$, $(2,1)_{1}$, $(1,2)_{1}$, $(2,2)_{1}$, and so on. Here $(L,0)_{L}$ is the so-called $\rho$-mode, and $(0,L)_{L}$ the so-called $\lambda$-mode.

Now we analyze the excitation energies of the $\lambda$-mode and the $\rho$-mode with $L=1$.
 Tables~\ref{t1} and~\ref{t2} in the appendix list the details of each term contribution to the eigenvalues of the Hamiltonian. For comparison, Types I, II and III in each table are presented, which correspond to the light-quark baryon, singly heavy baryon and doubly heavy baryon, respectively. $H$ in the last column is the total energy (the mass of the state). $\Delta$ in the brackets is the excitation energy of each term with respect to the ground state.

\begin{figure}[htbp]
\centering
\includegraphics[width=8.5cm]{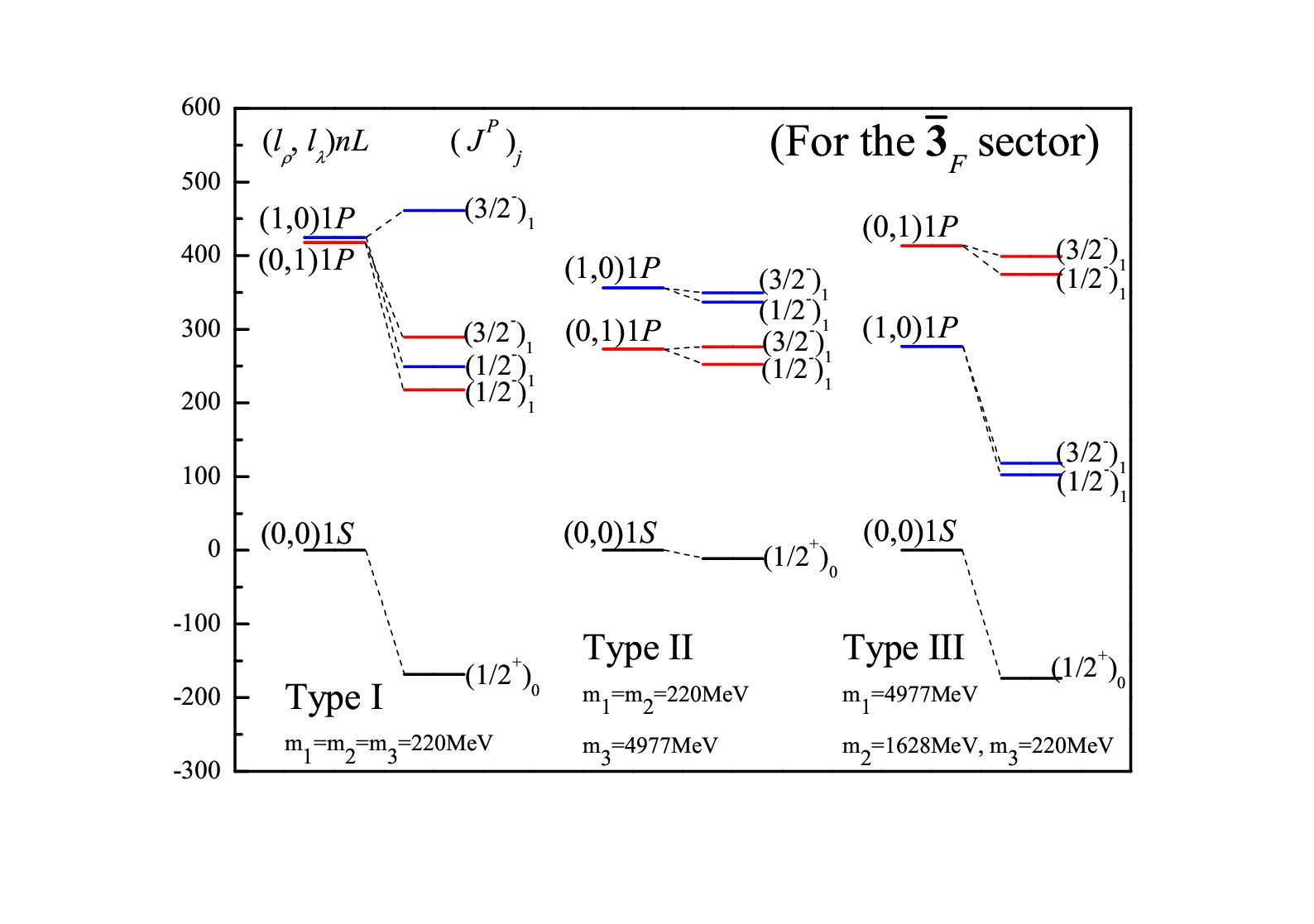}
\caption{Relative energy levels $\langle\Delta H_{mode}\rangle$ and the energy splittings of the $\rho$-mode and the $\lambda$-mode for the $\bar{\mathbf{3}}_{F}$ sector with Types I, II and III. The energy values are in MeV.}
\label{f01}
\end{figure}

\begin{figure}[htbp]
\centering
\includegraphics[width=8.5cm]{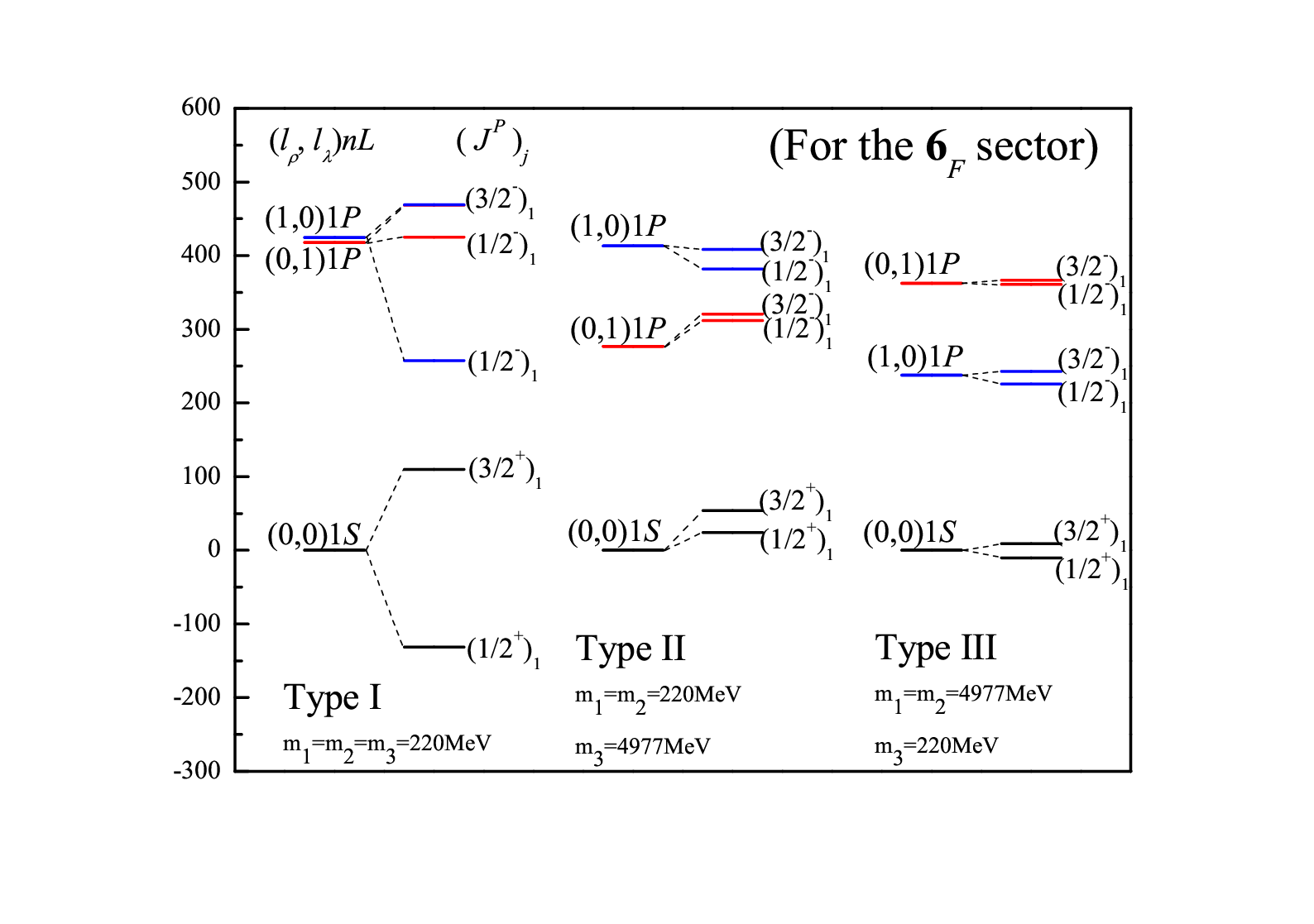}
\caption{Same as Fig.\ref{f01}, but for the $\textbf{6}_{F}$ sector.}
\label{f02}
\end{figure}

As shown in Tables~\ref{t1} and~\ref{t2}, the energy levels of each $(l_{\rho},l_{\lambda})nL$ mode are dominated by the sum of $H_{0}$, $\sum G_{ij}$ and $\sum S_{ij}$. While $H^{hyp}$ and $H^{so}$ are responsible for the splitting of the levels. This feature can be seen more clearly in Figs.\ref{f01} and \ref{f02}. For the convenience of the following discussion, we define
 \begin{eqnarray}
H_{mode}=H_{0}+\sum G_{ij}+\sum S_{ij},
\end{eqnarray}
whose eigenvalues stand for the energy levels of the excitation modes $(l_{\rho},l_{\lambda})_{L}$. Then, we use $\langle\Delta H_{mode}\rangle$ to represent the excitation energies (relative energy levels) of the modes with respect to the pure $S$-wave mode $(0,0)_{0}$.

Compared with Type I, as shown in Fig.\ref{f01}, $m_{3}$ in Type II is replaced by a heavy quark mass ($m_{b}=4977$ MeV), which leads to two results in Type II: (1) The excitation energy of the $(0,1)1P$ mode is suppressed; (2) The corresponding energy splitting is also reduced. As a whole, the energy level of the $\lambda$-mode and the energy splitting are both suppressed due to the mass of the heavy quark. The same is true in Type III, nevertheless, here the excitation energy of the $\rho$-mode is suppressed which is caused by the heavy quark pair. The similar phenomenon can be seen for the $\mathbf{6}_{F}$ sector as shown in Fig.\ref{f02}.
The conclusion in this work is the same as that of Refs.\cite{f06,f10}.
We can uniformly summarize that the lower orbital excitation energy is caused by the heavy quark(s) in singly and doubly heavy baryons. This is the HQD mechanism.

 This mechanism can be understood in a semi-classical way. The orbital excitation generates the rotational kinetic energy $E_{r.k.}$, which increases the kinetic energies of quarks and the radial distance between quarks. The former contributes directly to $H_{_{0}}$. While the latter raises the values of $\sum G_{ij}$ and $\sum S_{ij}$ due to the one-gluon exchange potential $G(r)=-\frac{4\alpha_{s}(r)}{3r}$ and linear confinement potential $S(r)=b'r+c'$. Here $b'$ and $c'$ are constants. $\alpha_{s}(r)$ is the running coupling constant and has been parameterized~\cite{f05}. Therefore, the rotational kinetic energy $E_{r.k.}$ determines $\langle\Delta H_{mode}\rangle$ finally. While the rotational kinetic energy in a given state $\langle E_{r.k.}\rangle$ depends on the moment of inertia of the state $\langle I\rangle$ through $\langle E_{r.k.}\rangle=\frac{\langle L^{2}\rangle}{2\langle I\rangle}$. So, a larger $\langle I\rangle$ will cause a smaller $\langle E_{r.k.}\rangle$, which will eventually lead to a lower $\langle\Delta H_{mode}\rangle$.

For a two-body system rotating in a plane, the following relationships exist, $m_{A}r_{A}=m_{B}r_{B}$ and $r_{A}+r_{B}=r$. Here $m_{A(B)}$ denotes the mass of $A(B)$ particle. $r_{A(B)}$ is the corresponding distance to the center of mass. Then, the moment of inertia of the system $I=m_{A}r_{A}^{2}+m_{B}r_{B}^{2}=\frac{m_{A}m_{B}}{m_{A}+m_{B}}r^{2}$.
For the $P$-wave orbital excitation of singly heavy baryons, the $\lambda$-mode means ($l_{\rho},l_{\lambda}$)=(0,1). If we take the baryon as a two-body system with $m_{A}=m_{Q}\gg m_{q}$ and $m_{B}=m_{q_{1}}+m_{q_{2}}\sim 2m_{q}$, $\langle I_{\lambda}\rangle \sim 2m_{q}\langle r_{\lambda}^{2}\rangle$ is obtained based on the above discussion. In the $\rho$-mode with ($l_{\rho},l_{\lambda}$)=(1,0), the rotation occurs in the two light quarks. And $\langle I_{\rho}\rangle \sim \frac{1}{2}m_{q}\langle r_{\rho}^{2}\rangle$ as $m_{A}\sim m_{B}\sim m_{q}$. In this way, for both of the $(\bar{\mathbf{3}}_{F})$ and $(\mathbf{6}_{F})$ sectors, we can obtain the moment of inertia $\langle I\rangle$ for Types I, II and III. By using the
calculated values of $\langle r_{\rho}^{2}\rangle^{1/2}$ and $\langle r_{\lambda}^{2}\rangle^{1/2}$, we can estimate the value of the rotational kinetic energy $\langle E_{r.k.}\rangle$ and compare it with the excitation energy $\langle\Delta H_{mode}\rangle$.

The used variables and comparison results are listed in Table~\ref{t2p}. It is shown that for Type II (the singly heavy baryons), the estimated values of $\langle E_{r.k.}\rangle$ in the $\lambda$-mode are lower than those in the $\rho$-mode, being the bigger moment of inertia $\langle I\rangle$ in the $\lambda$-mode due to the influence of the heavy quark. By comparison, it can be seen that the estimated $\langle E_{r.k.}\rangle$ is near $\langle\Delta H_{mode}\rangle$. For Type III (the doubly heavy baryons), we can conclude that the heavy quarks dominate the lower $\langle E_{r.k.}\rangle$ in the $\rho$-mode, which qualitatively agrees with $\langle\Delta H_{mode}\rangle$ as shown in Table~\ref{t2p}.

The above discussion confirms our understanding that the rotational kinetic energies $E_{r.k.}$ of the excitation modes involving heavy quark(s) are significantly lower, which determines the relative energy levels of the excitation modes. Therefore, the HQD mechanism has been understood in the semi-classical way.
On the other hand, the suppression of the energy levels splitting for heavy baryons can be explained in the heavy quark limit~\cite{f06,f13p,f13p2}. We may conclude that the excitation mode with lower energy levels is always associated with the heavy quark(s), and the energy level splitting is also suppressed by the heavy quark(s). So, the heavy quarks dominate the orbital excitation of singly and doubly heavy baryons.

\section{The test of HQD approximation in singly heavy baryons}\label{sec3}

\subsection{The orbital excitation of singly heavy baryons}\label{sec3.1}

\begin{figure*}[htbp]
\centering
\includegraphics[width=18cm]{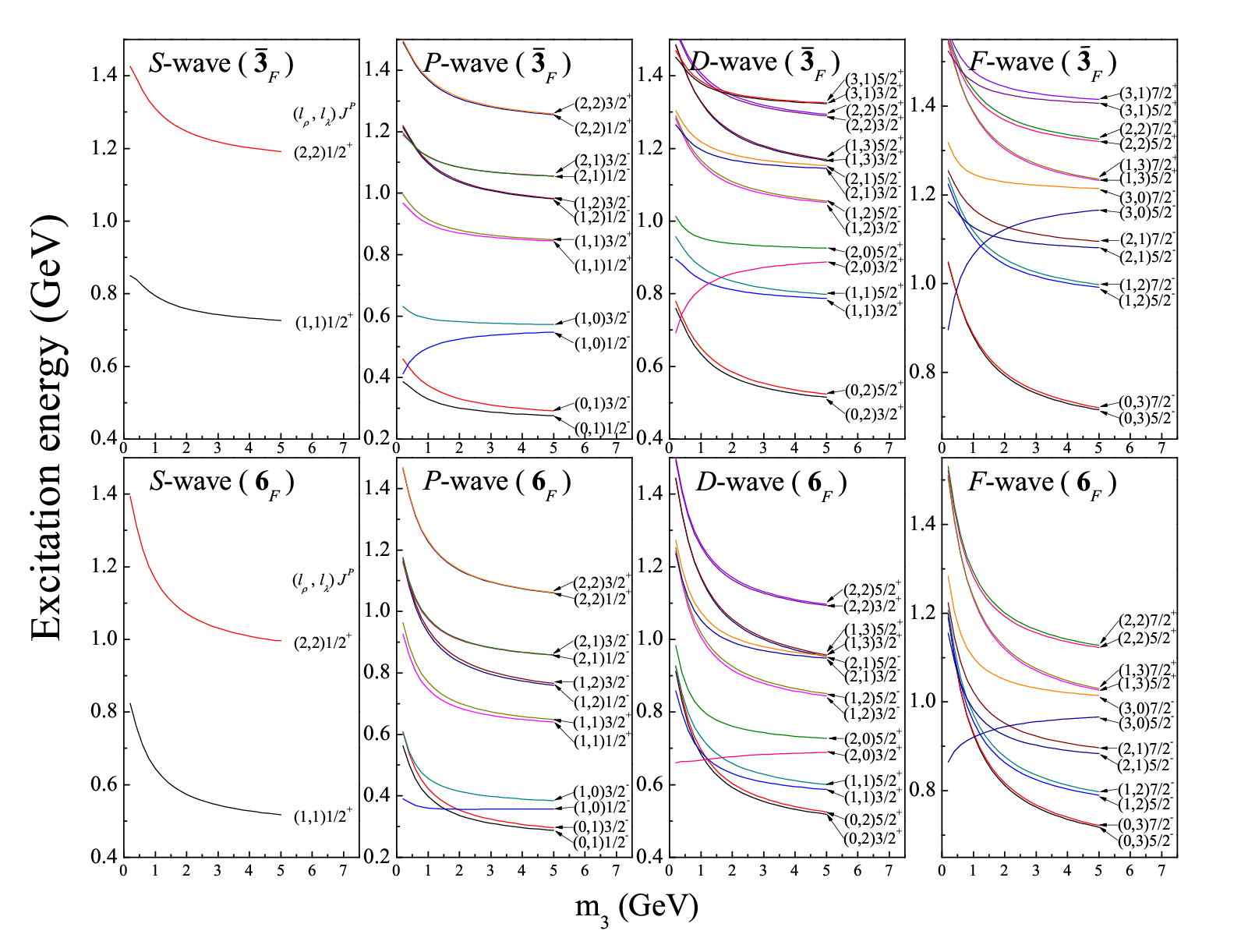}
\caption{The evolution of excitation energies with $m_{3}$ for different modes of singly heavy baryons. Here only the modes with $J=L\pm\frac{1}{2}$ are selected. $m_{3}$ changes from 0.2 GeV to 5.0 GeV. The excitation energies are measured from the pure $S$-wave states $1S(\frac{1}{2}^{+})_{0}$ for the $\bar{\mathbf{3}}_{F}$ and $1S(\frac{1}{2}^{+})_{1}$ for the $\mathbf{6}_{F}$, respectively. And $m_{1}=m_{2}=0.22$ GeV is used.}
\label{f2}
\end{figure*}

 For further investigating the HQD approximation in singly heavy baryons, the evolution of the excitation energies with $m_{3}$ for different modes is plotted in the cases of the $P$-, $D$- and $F$-wave states. At the same time, the relative energies of the $(1,1)1S(\frac{1}{2}^{+})$ and $(2,2)1S(\frac{1}{2}^{+})$ states with respect to the pure $S$-wave state $(0,0)1S(\frac{1}{2}^{+})$ are also given for comparison, as shown in Fig.\ref{f2}.

 In the $\bar{\mathbf{3}}_{F}$ sector, the energies of the $(1,1)1S(\frac{1}{2}^{+})$ and $(2,2)1S(\frac{1}{2}^{+})$ states are obviously higher than that of the pure $S$-wave state.
 For the $P$-, $D$- and $F$-wave states, the excitation energies of most modes are quickly suppressed as $m_{3}$ increases, except for the $\rho$-mode.
 And the $\lambda$-mode is indeed the one with the lowest excitation energy.

 It is reasonable to believe that the excitation mode with the lowest energy is most stable and has the greatest probability of being observed experimentally.
 Then, all excitation modes with higher energies can be approximately ignored. Thus, the `true' ground state comes from the pure $S$-wave state, and the `real' orbital excited state originates from the $\lambda$-mode, which is dominated by the heavy quark(s). This is the HQD approximation as mentioned before in this paper.

 The situation of the $\mathbf{6}_{F}$ sector is similar to that of the $\bar{\mathbf{3}}_{F}$ sector that the $\lambda$-mode is dominant. But the phenomenon is not so obvious when $m_{3}\sim$~1.6 GeV for the $P$-wave states, where the excitation energy of the $(0,1)1P(\frac{1}{2}^{+})_{1}$ state is only slightly lower than that of the $(1,0)1P(\frac{1}{2}^{+})_{1}$ state. It implies that $c$ quark ($m_{c}=1.628$GeV) is not heavy enough and hardly treated as a heavy quark, which was also pointed out in some theoretical works~\cite{f13}.
Even so, the HQD approximation is still applicable to the $\mathbf{6}_{F}$ sector.

\subsection{Predicted excited states with HQD approximation}\label{sec3.2}
With the HQD approximation, the excited spectra of singly heavy baryons were systemically investigated with a unified set of parameters~\cite{f11,f11p}. The results show that the predicted masses reproduce the experimental data well.
As shown in Table~\ref{t3}, for the well established baryons in experiment~\cite{f14,f14p}, the calculated masses are very close to the data, apart from $\Lambda_{c}(2940)^{+}$~\cite{f11,f15}. The maximum difference of the mass values between theory and experiment is generally no more than 20 MeV.

\begin{figure*}[htbp]
\centering
\includegraphics[width=16cm]{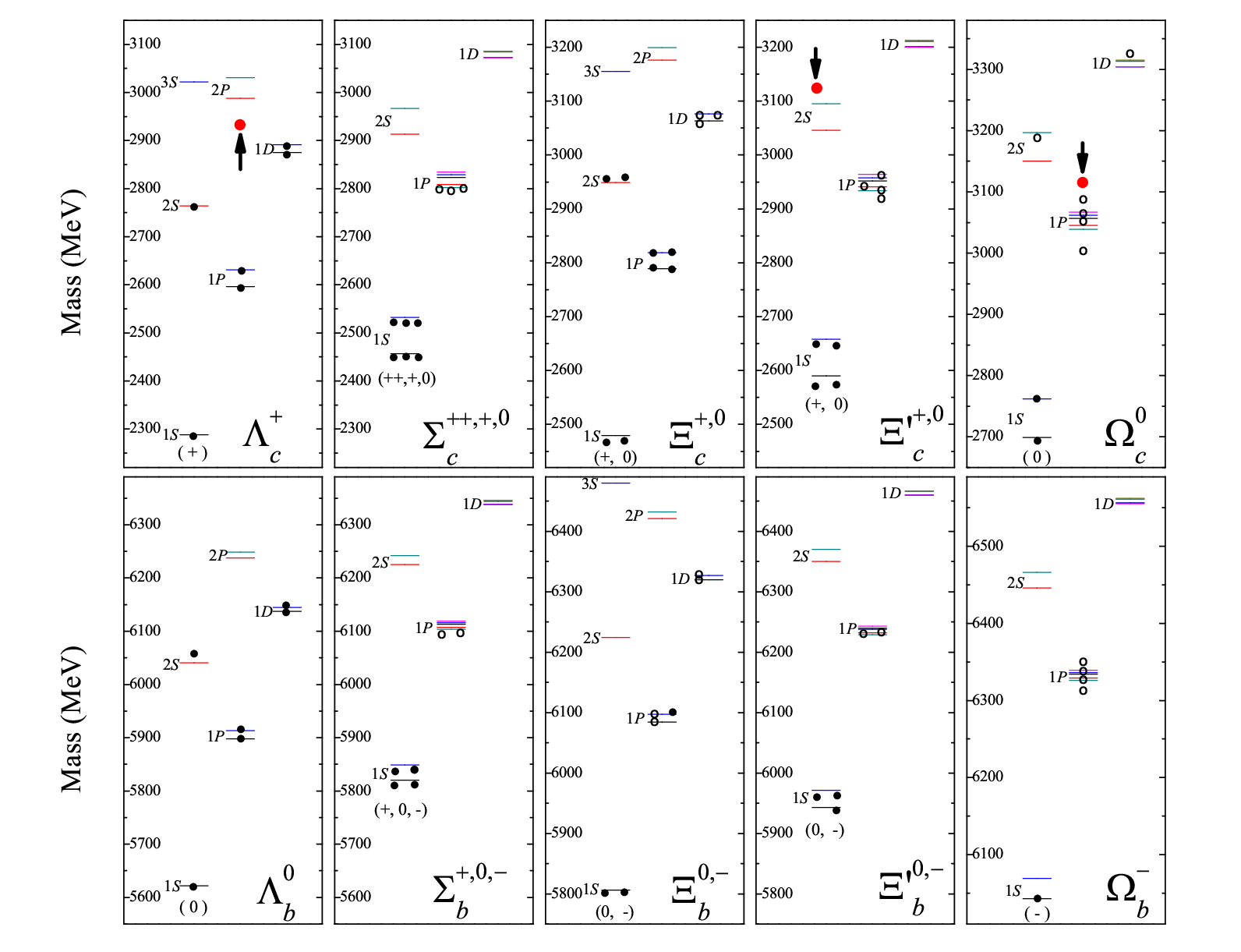}
\caption{Calculated spectra of singly heavy baryons and the relevant experimental data. `++',`+', `0' and `-' in the brackets indicate the charge states of baryons. The solid black circles denote the baryons in Table~\ref{t1}, and the open circles are the ones in Table~\ref{t2}. The three bigger red circles represent $\Lambda_{c}(2940)^{+}$, $\Xi_{c}(3123)^{+}$, and $\Omega_{c}(3120)^{+}$, respectively, which can not be assigned very well. More excited states can be found in Refs.~\cite{f11,f11p}.}
\label{f3}
\end{figure*}

For the observed baryons whose $J^{P}$ values are not yet confirmed~\cite{f14,f16,f16p1,f16p2,f16p3,f16p4}, the calculated results are also helpful for their rational assignments. As shown in Table~\ref{t4}, these baryons have been divided into several groups. In particular, some members in the same family were grouped together and labeled with $1P(\mathbf{6}_{F})$. With the help of the mass spectra, one can see them more clearly. As shown in Fig.\ref{f3}, for $\Sigma_{c(b)}$, $\Xi_{c(b)}^{'}$ and $\Omega_{c(b)}$, there are five excited states for the $1P$-wave, and their energies are almost degenerate. The five states are $1P(\frac{3}{2}^{-})_{2}$, $1P(\frac{1}{2}^{-})_{1}$, $1P(\frac{1}{2}^{-})_{0}$, $1P(\frac{3}{2}^{-})_{1}$ and $1P(\frac{5}{2}^{-})_{2}$, and increase in energy.
Some baryons of these $1P$ states have been observed as shown in Table~\ref{t4} and Fig.\ref{f3}. However, because their energies are so close to each other, the identification of these baryons becomes very difficult. So they can only be labeled with $1P(\mathbf{6}_{F})$, as shown in Table~\ref{t4}.

It is worth noting that the predicted spectra were obtained in 2022. After that, four baryons $\Omega_{c}(3185)^{0}$, $\Omega_{c}(3327)^{0}$~\cite{f16p3}, $\Xi_{b}(6095)^{0}$ and $\Xi_{b}(6087)^{0}$~\cite{f16p4} were observed by the LHCb collaboration. As expected, $\Omega_{c}(3185)^{0}$, $\Xi_{b}(6095)^{0}$ and $\Xi_{b}(6087)^{0}$ can be directly assigned to suitable states in Table~\ref{t4}. And $\Omega_{c}(3327)^{0}$~\cite{f16p3}, for the same reason mentioned above, should be labeled with $1D(\mathbf{6}_{F})$, where a total of six quasi-degenerate states are overlapped together (for details, see~\cite{f11}). In general, almost all baryons can be in good agreement with the predicted states. The exception is that $\Lambda_{c}(2940)^{+}$~\cite{f11,f15}, $\Xi_{c}(3123)^{+}$~\cite{f11p,f16p2} and $\Omega_{c}(3120)^{+}$~\cite{f14,f11} can not be assigned reasonably into the spectra of Fig.\ref{f3}, which needs further research and confirmation in theory and experiment .

Fig.\ref{f3} shows the predicted spectra agree well with the experimental data. And the `missing resonance' problem does not seem to exist here. It makes one more confident that the HQD mechanism is reasonable and the HQD approximation is feasible.
In addition, Fig.\ref{f3} shows an obvious shell structure of the spectrum. It implies that $L$ might be approximated as a good quantum number. If this is true, the $\gamma$ transitions would occur between the neighbouring orbits. This feature has been partially observed in experiment. For example, $\Xi_{c}(2815)^{0}\rightarrow \Xi_{c}^{0}\gamma$ is the $\gamma$ transition between the $1P(\frac{3}{2}^{-})$ and $1S(\frac{1}{2}^{+})$ states~\cite{f14}. Actually, the statement of `$L$-wave' has been adopted in many works such as Refs.~\cite{f17a0,f17,f17p0,f17p1,f17p2,f17p3,f17p4,f17p5,f17p6,f17p7,f17p8,f17p9,f17p10,f17p11,f17p12,f17p13,f17p14}. In this work, it is further identified.

\section{HQD mechanism in doubly heavy baryons}\label{sec4}

\begin{figure*}[htbp]
\centering
\includegraphics[width=16cm]{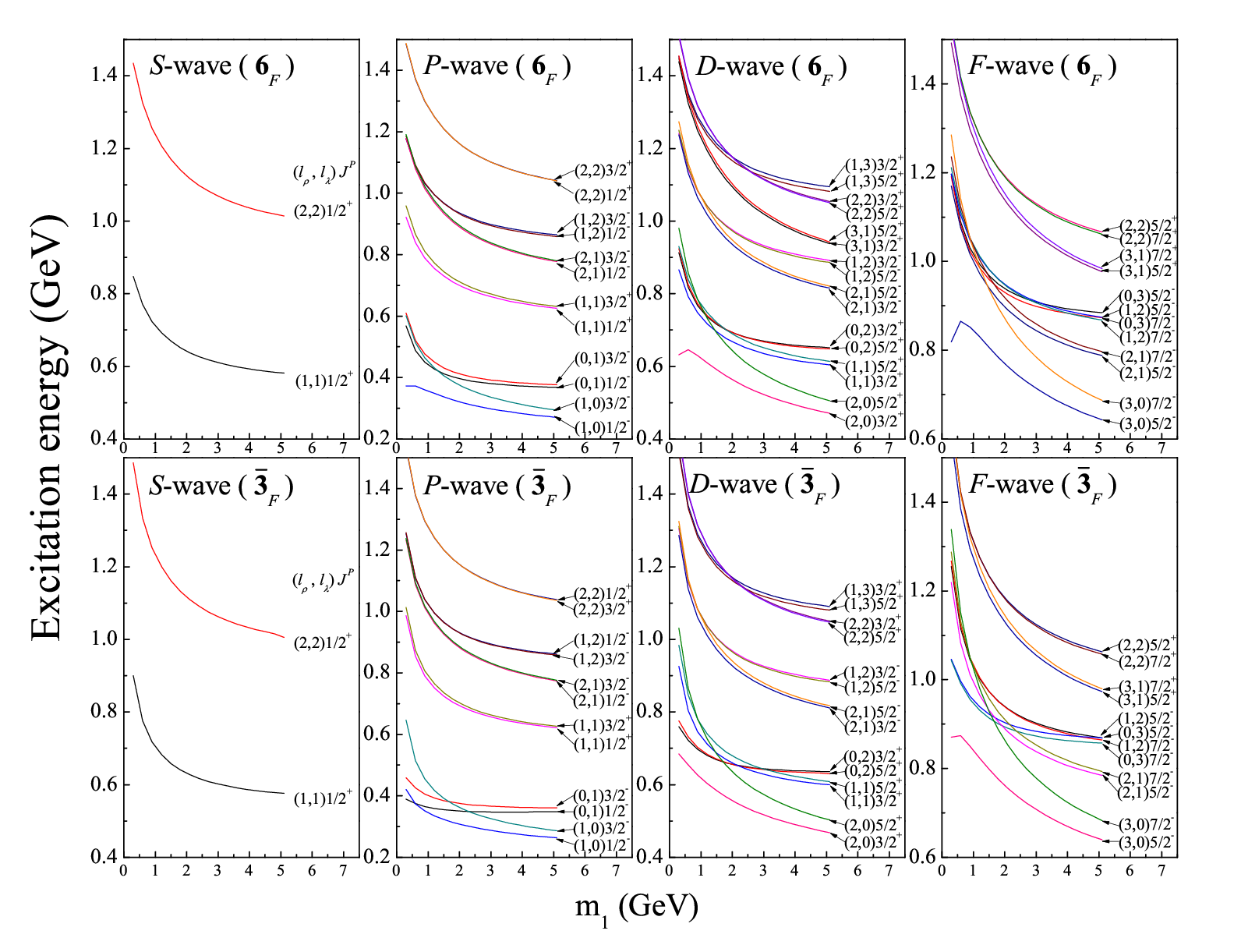}
\caption{Evolution of excitation energies with $m_{1}$ for different modes of doubly heavy baryons, where $m_{1}$ changes from 0.3 GeV to 5.1 GeV. The value of $m_{3}$ is equal to 0.22 GeV. $m_{2}$ is set as $m_{2}=m_{1}$ for the $\mathbf{6}_{F}$ sector. For the $\bar{\mathbf{3}}_{F}$ sector, $m_{2}$ is set as $m_{2}=\frac{m_{c}}{m_{b}}\times m_{1}$, which corresponds to $\Xi_{bc}$ or $\Omega_{bc}$ when $m_{1}=4.977$ GeV.}
\label{f4}
\end{figure*}

 The doubly heavy baryons $\Xi_{cc}$($\Omega_{cc}$), $\Xi_{bb}$($\Omega_{bb}$) and $\Xi'_{bc}$($\Omega'_{bc}$) belong to the $\mathbf{6}_{F}$ sector, and $\Xi_{bc}$($\Omega_{bc}$) belongs to the $\bar{\mathbf{3}}_{F}$ sector, as mentioned in Sect.\ref{sec2} A. To examine the HQD mechanism in the $\mathbf{6}_{F}$ sector, we set $m_{2}=m_{1}$ and change $m_{1}$ from 0.3 GeV to 5.1 GeV. For the $\bar{\mathbf{3}}_{F}$ sector, $m_{2}$ is set to be $\frac{m_{c}}{m_{b}}\times m_{1}$, which stands for $\Xi_{bc}$ (or $\Omega_{bc}$) baryon when $m_{1}$ approaches to 5.0 GeV (see the caption in Fig.\ref{f4}).

 As shown in Fig.\ref{f4}, the orbital excitation of doubly heavy baryons presents the same feature as that of singly heavy baryons. In this case, nevertheless, the mode with the lowest energy is the $\rho$-mode and the lowest orbital excitation energy is still dominated by the heavy quarks. Hence, the HQD mechanism remains valid and the HQD approximation is feasible for doubly heavy baryons.

\begin{figure*}[htbp]
\centering
\includegraphics[width=16cm]{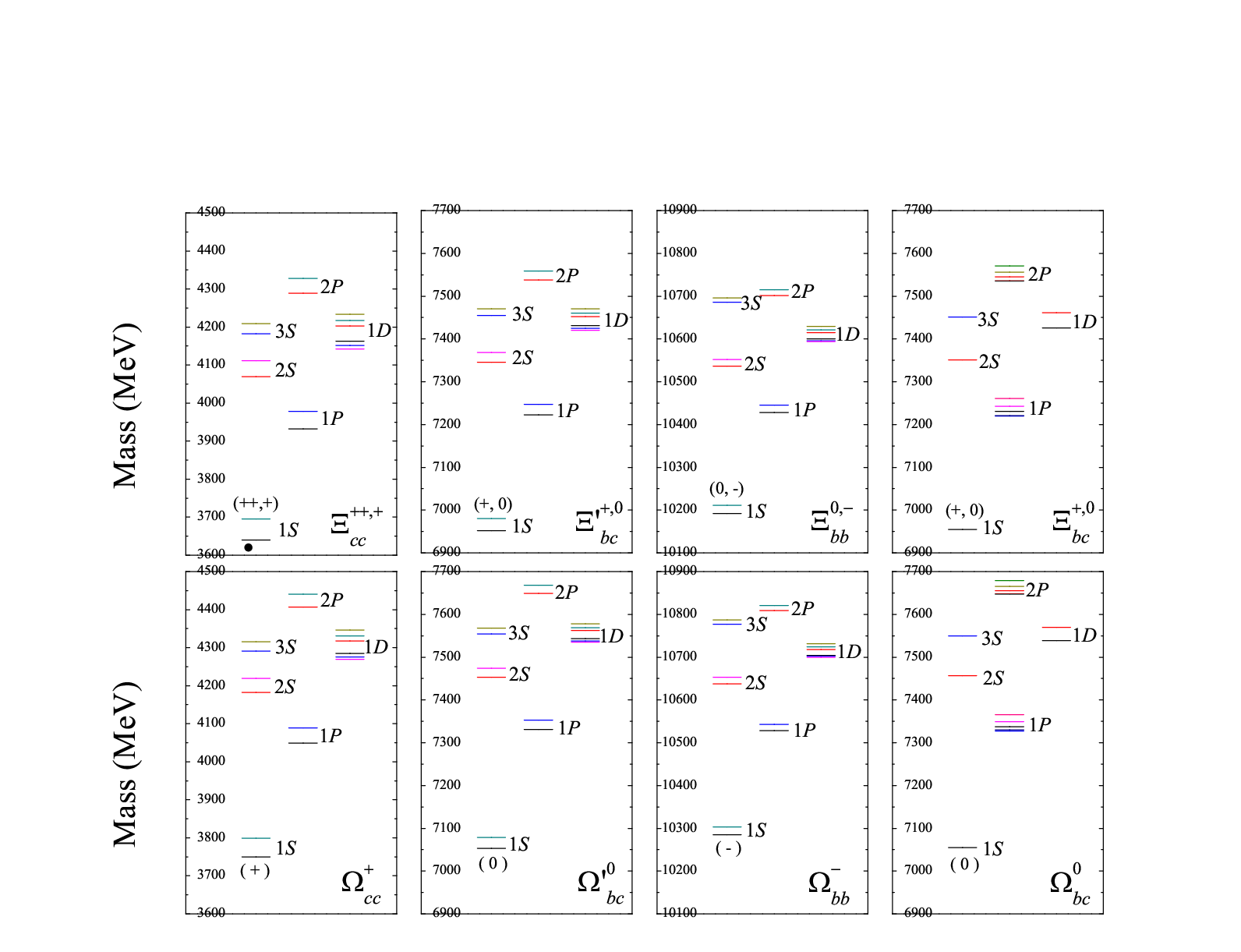}
\caption{Same as Fig.\ref{f3}, but for doubly heavy baryons. The solid black circle denotes the observed baryon $\Xi_{cc}^{++}$. More excited states can be found in Refs.~\cite{f12,f12p,f12p2}.}
\label{f5}
\end{figure*}

 Based on the HQD approximation, the spectra of doubly heavy baryons were investigated in the $\rho$-mode~\cite{f12,f12p,f12p2}. The predicted ground state mass of $\Xi_{cc}$ is 3640 MeV, which is very close to the experimental data 3621.6$\pm$0.4 MeV~\cite{f14}. The spectrum also presents an obvious shell structure as shown in Fig.\ref{f5}. It is expected that more doubly heavy baryons will be observed and can be used to further test the HQD approximation.

In addition, some works have noticed the excitation of the heavy quark pair is important and can not be ignored\cite{f09p1,f09p3}. In those works, the spectra of
the doubly heavy baryons therefore contain the excitation of the heavy quark pair. And the heavy quark pair was not treated as a point-like diquark, which was used in some other works\cite{f18p,f18p2,f18p3}. Unlike both of them, the HQD approximation freezes the orbital excitation of the $\lambda$-mode and only considers the $\rho$-mode, which predicts fewer excited states than the former\cite{f09p1,f09p3}, and lowers the excited energy levels than the latter\cite{f18p,f18p2,f18p3}.

\section{Signature of HQD mechanism in strong decay}\label{sec5}

If the HQD mechanism is reasonable, an interesting phenomenon would occur in the strong decay. The calculation shows that for a singly heavy baryon, the root-mean-square value of $r_{\lambda}$ increases with the orbital excitation~\cite{f11,f11p}. It means the heavy quark is further away from the light quark pair, as the orbital quantum number increases. For a high-orbital excited state, when the strong decay occurs, the heavy quark is therefore likely to capture the light antiquark which comes from the creation of a quark-antiquark pair in the vacuum, according to the $^{3}P_{0}$ model~\cite{f20,f20p1}. So, the heavy-light meson is more likely to be produced.

From the Particle Data Group (PDG)~\cite{f14}, one can see the $D^{0}p$ decay mode occurs in $\Lambda_{c}(2860)^{+}$ and $\Lambda_{c}(2880)^{+}$, which are identified as the $D$-wave states in theory~\cite{f07p,f11,f19}.
Similarly, the $\Lambda D^{+}$ decay modes are also observed experimentally for $\Xi_{c}(3055)^{+}$ and $\Xi_{c}(3080)$. So, they are assigned to the $1D(\frac{3}{2}^{+},\frac{5}{2}^{+})$ doublet states of the $\bar{\mathbf{3}}_{F}$ sector in Table~\ref{t4}. For $\Lambda_{c}^{+}$, the ground state of the $\Lambda_{c}$ family, the exotic mode of $p\gamma D$ has been observed in experiment. But its branching ratio is very small. Additionally, the $pD^{0}$ mode is observed in the strong decay of $\Lambda_{c}(2940)^{+}$, which means it should be a high-orbital excited state.

 In Table~\ref{t4} and Fig.\ref{f3}, $\Xi_{b}(6327)^{0}$ and $\Xi_{b}(6333)^{0}$ is identified as the $D$-wave doublet states in this work. In their observed strong decay modes, however, the heavy-light meson is absent~\cite{f14}. From Ref.~\cite{f11p}, we find their $\langle r_{\lambda}^{2}\rangle^{1/2}$ values are 0.743 fm and 0.751 fm, respectively, which are smaller than those (0.825 fm and 0.843 fm) of the $D$-wave states $\Xi_{c}(3055)^{+}$ and $\Xi_{c}(3080)$. It implies the heavy-light meson is more difficult to be produced in the strong decay of the bottom baryons, as compared to the charmed baryons.

So far, the two-body strong decay mode containing a heavy-light meson is mainly observed in the high-orbital excited states of the charmed baryons. The same phenomenon is expected to occur in doubly heavy baryons, which might be helpful in searching for the doubly heavy baryons with the high-orbital excitation in experiments.

\begin{large}
\section{Conclusions}\label{sec6}
\end{large}

The singly and doubly heavy baryons are special, due to the large mass values of charm and bottom quarks. Since these heavy quarks are much more massive than the light quarks, they break the mass symmetry. So, the $SU(3)_{F}$ flavor symmetry valid for light-quark baryons are not well applied here. On the other hand, the heavy quark limit is not fully applicable here either, because the masses of the heavy quarks are much less than infinity. Inspired by the progress of the heavy baryon spectroscopy in recent decades, we speculate a mechanism of the heavy quark dominance (HQD) in the orbital excitation of singly and doubly heavy baryons.

By analyzing the Hamiltonian of the relativistic quark model for a three-quark system, we tried to understand the HQD mechanism. It is shown that the excitation mode with lower energy levels is always associated with the heavy quark(s), and the splitting of the energy levels is suppressed by the heavy quark(s) as well. In other words, the heavy quarks dominate the orbital excitation of singly and doubly heavy baryons. In a semi-classical way, we show that the rotational kinetic energy generated by the orbital excitation is positively related to the excitation energy of the mode. And the smaller rotational kinetic energy is determined by the heavy quark(s). So, the heavy quarks eventually dominate the lower orbital excitation for singly and doubly heavy baryons. Thus, we get a physical understanding of this mechanism.

The HQD mechanism is then tested by comparing various orbital excitation modes for singly heavy baryons. It is shown that the excitation mode dominated by the heavy quark(s) has lower energy levels, therefore, it is most stable and has the greatest probability of being observed experimentally, which justifies the HQD approximation.
 With the HQD approximation, the predicted excited spectra of singly heavy baryons can reproduce the available experimental data nicely. Additionally, the shell structure of the mass spectra implies $L$ can be regarded as a good quantum number approximately, which is supported by the $\gamma$ transition in experiment. Thus, the HQD mechanism is proved to be reasonable in singly heavy baryons. For doubly heavy baryons, their orbital excitations are also dominated by the heavy quarks. The corresponding mass spectra with the HQD approximation are expected to be tested in future experiments.
The HQD mechanism also predicts an interesting phenomenon that the heavy-light meson is more likely to be produced in the strong decay of the high-orbital excited states for singly and doubly heavy baryons. This kind of strong decay mode may be the signature of the HQD mechanism as well.

 The HQD mechanism is rooted in the breakdown of the mass symmetry in a multi-quark system. Therefore, it should be valid not only for singly and doubly baryons, but also for the multi-quark systems containing one or two heavy quarks, such as the tetraquarks $Qqqq$ and $QQqq$, or the pentaquarks $Qqqqq$ and $QQqqq$.

\begin{large}
\section*{Acknowledgements}
\end{large}

We thank the reviewers of our series of work on the heavy baryon spectroscopy, for their helpful comments, questions and suggestions, which prompted our consideration of this work. This research was supported by the National Natural Science Foundation of China (Grant Nos. 11675265, 12175068), the Continuous Basic Scientific Research Project (Grant No. WDJC-2019-13), and the Leading Innovation Project (Grant No. LC 192209000701).

\section*{Appendix}

\begin{table*}[htbp]
\begin{ruledtabular}\caption{Contribution of each term to the eigenvalues of $H$ for the states of the $(0,0)_{0}$, $(0,1)_{1}$ and $(1,0)_{1}$ modes of the $\bar{\mathbf{3}}_{F}$ sector (all quantities are in MeV).}
\label{t1}
\begin{tabular}{c c c c c c c c c c c c c}
$(l_{\rho},l_{\lambda})nL(J^{P})_{j}$ & $H_{_{0}}~(\Delta)$ & $G_{12}~(\Delta)$ & $G_{13(23)}~(\Delta)$ & $\sum G_{ij}~(\Delta)$ & $S_{12}~(\Delta)$ & $S_{13(23)}~(\Delta)$ & $\sum S_{ij}~(\Delta)$ &$H_{mode}~(\Delta)$& $H^{hyp}~(\Delta)$ & $H^{so}~(\Delta)$ & $H~(\Delta)$\\ \hline
\multicolumn{11}{c}{Type I ($m_{1}=m_{2}=m_{3}=220$ MeV)}\\
$(0,0)1S(\frac{1}{2}^{+})_{0}$ & 737 (0)  & -45 (0) & -48 (0) & -141 (0)  & 245 (0) & 152 (0) &  549 (0)  & 1145 (0)  & -168 (0)   & 0 (0)      & 978 (0)   \\
$(0,1)1P(\frac{1}{2}^{-})_{1}$ & 762 (25) & -45 (0) & -31 (17) & -107 (34) & 299 (54) & 305 (153)&  909 (360)&1564 (419) & -168 (0)& -53 (-53)& 1364 (386)  \\
$(0,1)1P(\frac{3}{2}^{-})_{1}$ & 762 (25) & -45 (0) & -31 (17) & -107 (34) & 299 (54) & 305 (153)&  909 (360)&1564 (419) & -141 (27)& 24 (24)& 1435 (457) \\
$(1,0)1P(\frac{1}{2}^{-})_{1}$ & 770 (33) & -31 (14) & -38 (10) & -108 (33) & 499 (254)& 205 (53)&  909 (360)& 1571 (426)& -148 (20)& -52 (-52)& 1395 (417) \\
$(1,0)1P(\frac{3}{2}^{-})_{1}$ & 770 (33) & -31 (14) & -38 (10) & -108 (33) & 499 (254)& 205 (53)&  909 (360)&1571 (426)&58 (226)  & -47 (-47)& 1607 (629) \\\hline
\multicolumn{11}{c}{Type II ($m_{1}=m_{2}=220$ MeV and $m_{3}=4977$ MeV)}\\
$(0,0)1S(\frac{1}{2}^{+})_{0}$ & 5473 (0)  & -45 (0) & -48 (0)   & -141 (0) & 239 (0) & 112 (0) &  463 (0)&5795 (0) & -173 (0) & 0 (0) & 5622 (0)   \\
$(0,1)1P(\frac{1}{2}^{-})_{1}$ & 5482 (9)  & -45 (0) & -28 (20)  & -102 (39) & 290 (51) & 201 (89)&  692 (229)&6072 (277)&-167 (6)& -7 (-7)& 5898 (276)  \\
$(0,1)1P(\frac{3}{2}^{-})_{1}$ & 5482 (9)  & -45 (0) & -28 (20)  & -102 (39) & 290 (51) & 201 (89)&  692 (229)&6072 (277)&-162 (11)& 4 (4)& 5913 (291) \\
$(1,0)1P(\frac{1}{2}^{-})_{1}$ & 5505 (32) & -31 (14) & -36 (12)  & -103 (38) & 491 (252)& 157 (45)&  806 (343)&6208 (413)&-14 (159)& -50 (-50)& 6170 (548) \\
$(1,0)1P(\frac{3}{2}^{-})_{1}$ & 5505 (32) & -31 (14) & -36 (12)  & -103 (38) & 491 (252)& 157 (45)&  806 (343)&6208 (413)&10 (183)& -49 (-49)& 6194 (572) \\\hline
\multicolumn{11}{c}{Type III ($m_{1}=4977$ MeV, $m_{2}=1628$ MeV and $m_{3}=220$ MeV)}\\
$(0,0)1S(\frac{1}{2}^{+})_{0}$ & 6853 (0)  & -88 (0)& -55 (0)  & -199 (0)   & 66 (0) & 124 (0) &  313 (0) &6967 (0)&-12 (0) & 0 (0) & 6955 (0)   \\
$(0,1)1P(\frac{1}{2}^{-})_{1}$ & 6870 (17)  & -88 (0)& -27 (28) & -143 (56) & 74 (8) & 261 (137)&  596 (283)&7323 (356)&-13 (-1)& -26 (-26)& 7304 (349)  \\
$(0,1)1P(\frac{3}{2}^{-})_{1}$ & 6870 (17)  & -88 (0)& -27 (28) & -143 (56) & 74 (8) & 261 (137)&  596 (283)&7323 (356)&-9 (3)  & 13 (13) & 7317 (362) \\
$(1,0)1P(\frac{1}{2}^{-})_{1}$ & 6856 (3) & -33 (55) & -43 (12)& -120 (79) & 192 (126)& 156 (32)&  504 (191)&7240 (273)&-16 (-4)& -7 (-7)& 7220 (265) \\
$(1,0)1P(\frac{3}{2}^{-})_{1}$ & 6856 (3) & -33 (55) & -43 (12)& -120 (79) & 192 (126)& 156 (32)&  504 (191)&7240 (273)&8 (20) & -7 (-7)& 7243 (288) \\
\end{tabular}
\end{ruledtabular}
\end{table*}

\begin{table*}[htbp]
\begin{ruledtabular}\caption{Contribution of each term to the eigenvalues of $H$ for the states of the $(0,0)_{0}$, $(0,1)_{1}$ and $(1,0)_{1}$ modes of the $\textbf{6}_{F}$ sector (all quantities are in MeV).}
\label{t2}
\begin{tabular}{c c c c c c c c c c c c c}
$(l_{\rho},l_{\lambda})nL(J^{P})_{j}$ & $H_{0}~(\Delta)$ & $G_{12}~(\Delta)$ & $G_{13(23)}~(\Delta)$ & $\sum G_{ij}~(\Delta)$ & $S_{12}~(\Delta)$ & $S_{13(23)}~(\Delta)$ & $\sum S_{ij}~(\Delta)$ &$H_{mode}~(\Delta)$& $H^{hyp}~(\Delta)$ & $H^{so}~(\Delta)$ & $H~(\Delta)$\\ \hline
\multicolumn{11}{c}{Type I ($m_{1}=m_{2}=m_{3}=220$ MeV)}\\
$(0,0)1S(\frac{1}{2}^{+})_{1}$ & 737 (0)  & -45 (0) & -48 (0) & -141 (0) & 245 (0) & 152 (0) &  549 (0)&1145 (0) & -131 (0) & 0 (0) & 1015 (0)   \\
$(0,0)1S(\frac{3}{2}^{+})_{1}$ & 737 (0)  & -45 (0) & -48 (0) & -141 (0) & 245 (0) & 152 (0) &  549 (0)&1145 (0)& 110 (241) & 0 (0) & 1256 (241)  \\
$(0,1)1P(\frac{1}{2}^{-})_{1}$ & 762 (25) & -45 (0) & -31 (17) & -107 (34) & 299 (54) & 305 (153)&  909 (360)&1564 (419)&21 (152)& -25 (-25)& 1571 (556)  \\
$(0,1)1P(\frac{3}{2}^{-})_{1}$ & 762 (25) & -45 (0) & -31 (17) & -107 (34) & 299 (54) & 305 (153)&  909 (360) &1564 (419)&45 (176)& 12 (12)& 1615 (600) \\
$(1,0)1P(\frac{1}{2}^{-})_{1}$ & 770 (33) & -31 (14) & -38 (10) & -108 (33) & 499 (254)& 205 (53)&  909 (360)&1571 (426)&-167 (-36)& $\sim$0 (0)& 1404 (389) \\
$(1,0)1P(\frac{3}{2}^{-})_{1}$ & 770 (33) & -31 (14) & -38 (10) & -108 (33) & 499 (254)& 205 (53)&  909 (360)& 1571 (426)&45 (176)& $\sim$0 (0)& 1615 (600) \\\hline
\multicolumn{11}{c}{Type II ($m_{1}=m_{2}=220$ MeV and $m_{3}=4977$ MeV)}\\
$(0,0)1S(\frac{1}{2}^{+})_{1}$ & 5473 (0)  & -45 (0) & -48 (0) & -141 (0) & 239 (0) & 112 (0) &  463 (0) &5795 (0)&24 (0) & 0 (0) & 5820 (0)   \\
$(0,0)1S(\frac{3}{2}^{+})_{1}$ & 5473 (0)  & -45 (0) & -48 (0)  & -141 (0) & 239 (0) & 112 (0) &  463 (0) & 5795 (0)&54 (30) & 0 (0) & 5849 (29)  \\
$(0,1)1P(\frac{1}{2}^{-})_{1}$ & 5482 (9)  & -45 (0) & -28 (20)  & -102 (39) & 290 (51) & 201 (89)&  692 (229)&6072 (277)&38 (14)& -3 (-3)& 6107 (287)  \\
$(0,1)1P(\frac{3}{2}^{-})_{1}$ & 5482 (9)  & -45 (0) & -28 (20)  & -102 (39) & 290 (51) & 201 (89)&  692 (229)& 6072 (277)&43 (19)& 1 (1)& 6116 (296) \\
$(1,0)1P(\frac{1}{2}^{-})_{1}$ & 5505 (32) & -33 (12) & -36 (12)  & -103 (38) & 491 (252)& 157 (45)&  805 (342)&6207 (412)&-31 (-55)& $\sim$0 (0)& 6178 (358) \\
$(1,0)1P(\frac{3}{2}^{-})_{1}$ & 5505 (32) & -33 (12) & -36 (12)  & -103 (38) & 491 (252)& 157 (45)&  805 (342)& 6207 (412)&-5 (-29)& $\sim$0 (0)& 6204 (384) \\\hline
\multicolumn{11}{c}{Type III ($m_{1}=m_{2}=4977$ MeV and $m_{3}=220$ MeV)}\\
$(0,0)1S(\frac{1}{2}^{+})_{1}$ & 10199 (0)  & -137 (0)& -56 (0)  & -250 (0)   & 18 (0) & 118 (0) &  254 (0) &10203 (0)&-10 (0) & 0 (0) & 10192 (0)   \\
$(0,0)1S(\frac{3}{2}^{+})_{1}$ & 10199 (0)  & -137 (0)& -56 (0) & -250 (0)  & 18 (0) & 118 (0) &  254 (0) &10203 (0)&9 (19) & 0 (0) & 10211 (19)  \\
$(0,1)1P(\frac{1}{2}^{-})_{1}$ & 10216 (17)  & -137 (0)& -27 (29) & -191 (59) & 21 (3) & 259 (141)&  539 (285)&10564 (361)&$\sim$0 (10)& -2 (-2)& 10563 (371)  \\
$(0,1)1P(\frac{3}{2}^{-})_{1}$ & 10216 (17)  & -137 (0)& -27 (29) & -191 (59) & 21 (3) & 259 (141)&  539 (285)&10564 (361)&3 (13)       & 1 (1) & 10569 (377) \\
$(1,0)1P(\frac{1}{2}^{-})_{1}$ & 10201 (2) & -53 (84) & -50 (6)& -152 (98) & 115 (97)& 138 (20)&  391 (137)&10440 (237)&-12 (-2)& $\sim$0 (0)& 10428 (236) \\
$(1,0)1P(\frac{3}{2}^{-})_{1}$ & 10201 (2) & -53 (84) & -50 (6)& -152 (98) & 115 (97)& 138 (20)&  391 (137)&10440 (237)&5 (15)& $\sim$0 (0)& 10445 (253) \\
\end{tabular}
\end{ruledtabular}
\end{table*}

\begin{table*}[htbp]
\begin{ruledtabular}\caption{Calculated $\langle r_{\rho}^{2}\rangle^{1/2}$ and $\langle r_{\lambda}^{2}\rangle^{1/2}$ (in fm) for the states of the $(0,0)_{0}$, $(0,1)_{1}$ and $(1,0)_{1}$ modes. And comparison of the rotational kinetic energy $\langle E_{r.k.}\rangle$ and the excitation energy $\langle\Delta H_{mode}\rangle$ (in MeV) of the above modes. The data in boldface are used in calculations ($m_{Q}=4977$ MeV and $m_{q}=220$ MeV). }
\label{t2p}
\begin{tabular}{c c c c c c | c c c c c c c c c c}
\multicolumn{6}{c |}{the $\bar{\mathbf{3}}_{F}$ sector} & \multicolumn{6}{c}{the $\textbf{6}_{F}$ sector}\\
$(l_{\rho},l_{\lambda})nL(J^{P})_{j}$ & $\langle r_{\rho}^{2}\rangle^{1/2}$ & $\langle r_{\lambda}^{2}\rangle^{1/2}$ & $\langle I \rangle$ & $\langle E_{r.k.}\rangle$ & $\langle\Delta H_{mode}\rangle$ & $(l_{\rho},l_{\lambda})nL(J^{P})_{j}$ & $\langle r_{\rho}^{2}\rangle^{1/2}$  & $\langle r_{\lambda}^{2}\rangle^{1/2}$  & $\langle I \rangle$ & $\langle E_{r.k.}\rangle$  & $\langle\Delta H_{mode}\rangle$ \\ \hline
\multicolumn{6}{c |}{Type I ($m_{1}=m_{2}=m_{3}=220$ MeV)} & \multicolumn{6}{c}{Type I ($m_{1}=m_{2}=m_{3}=220$ MeV)}\\
$(0,0)1S(\frac{1}{2}^{+})_{0}$ & 0.534  & 0.520 & - & - & -& $(0,0)1S(\frac{1}{2}^{+})_{1}$ & 0.609 & 0.471   & - & -& -\\
                 -          &  -        &  -   & -& -   &-& $(0,0)1S(\frac{3}{2}^{+})_{1}$ & 0.672 & 0.595 & - & - & -\\
$(0,1)1P(\frac{1}{2}^{-})_{1}$ & 0.566 & $\mathbf{0.717}$ & $\frac{2}{3}m_{q}\langle r_{\lambda}^{2}\rangle$ & $\sim516$ & 419 & $(0,1)1P(\frac{1}{2}^{-})_{1}$ & 0.701 & $\mathbf{0.770}$ & $\frac{2}{3}m_{q}\langle r_{\lambda}^{2}\rangle$ & $\sim447$ &419\\
$(0,1)1P(\frac{3}{2}^{-})_{1}$ & 0.587 & $\mathbf{0.798}$ & $\frac{2}{3}m_{q}\langle r_{\lambda}^{2}\rangle$ & $\sim414$ & 419 & $(0,1)1P(\frac{3}{2}^{-})_{1}$ & 0.718 & $\mathbf{0.823}$  & $\frac{2}{3}m_{q}\langle r_{\lambda}^{2}\rangle$ & $\sim390$ & 419\\
$(1,0)1P(\frac{1}{2}^{-})_{1}$ & $\mathbf{0.827}$ & 0.500 & $\frac{1}{2}m_{q}\langle r_{\rho}^{2}\rangle$ & $\sim516$ & 426& $(1,0)1P(\frac{1}{2}^{-})_{1}$ & $\mathbf{0.842}$ & 0.503 & $\frac{1}{2}m_{q}\langle r_{\rho}^{2}\rangle$ &$\sim498$ &426 \\
$(1,0)1P(\frac{3}{2}^{-})_{1}$ & $\mathbf{0.900}$ & 0.628 & $\frac{1}{2}m_{q}\langle r_{\rho}^{2}\rangle$ &$\sim436$ & 426 & $(1,0)1P(\frac{3}{2}^{-})_{1}$ &  $\mathbf{0.920}$ & 0.633  & $\frac{1}{2}m_{q}\langle r_{\rho}^{2}\rangle$ & $\sim417$ &426\\\hline
\multicolumn{6}{c |}{Type II ($m_{1}=m_{2}=220$ MeV, $m_{3}=4977$ MeV)} & \multicolumn{6}{c}{Type II ($m_{1}=m_{2}=220$ MeV, $m_{3}=4977$ MeV)}\\
$(0,0)1S(\frac{1}{2}^{+})_{0}$ & 0.519  & 0.407 & - & - &-& $(0,0)1S(\frac{1}{2}^{+})_{1}$ & 0.633 & 0.430  & - & -  &- \\
      -                     &-           & -     &-&  - &-& $(0,0)1S(\frac{3}{2}^{+})_{1}$ & 0.644  & 0.451 & - & - & -\\
$(0,1)1P(\frac{1}{2}^{-})_{1}$ & 0.534  & $\mathbf{0.577}$ & $2m_{q}\langle r_{\lambda}^{2}\rangle$ & $\sim265$ & 277& $(0,1)1P(\frac{1}{2}^{-})_{1}$ & 0.666  & $\mathbf{0.621}$ & $2m_{q}\langle r_{\lambda}^{2}\rangle$ & $\sim229$ & 277\\
$(0,1)1P(\frac{3}{2}^{-})_{1}$ & 0.539  & $\mathbf{0.590}$ & $2m_{q}\langle r_{\lambda}^{2}\rangle$ & $\sim253$ & 277& $(0,1)1P(\frac{3}{2}^{-})_{1}$ & 0.670  & $\mathbf{0.630}$ & $2m_{q}\langle r_{\lambda}^{2}\rangle$ & $\sim222$ &277\\
$(1,0)1P(\frac{1}{2}^{-})_{1}$ & $\mathbf{0.852}$ & 0.440 & $\frac{1}{2}m_{q}\langle r_{\rho}^{2}\rangle$ & $\sim486$ & 413 & $(1,0)1P(\frac{1}{2}^{-})_{1}$ & $\mathbf{0.869}$  & 0.444 & $\frac{1}{2}m_{q}\langle r_{\rho}^{2}\rangle$ & $\sim467$ & 412\\
$(1,0)1P(\frac{3}{2}^{-})_{1}$ & $\mathbf{0.865}$ & 0.462 & $\frac{1}{2}m_{q}\langle r_{\rho}^{2}\rangle$ & $\sim472$  &413 & $(1,0)1P(\frac{3}{2}^{-})_{1}$ & $\mathbf{0.883}$  & 0.466 & $\frac{1}{2}m_{q}\langle r_{\rho}^{2}\rangle$ & $\sim452$ & 412\\\hline
\multicolumn{6}{c |}{Type III ($m_{1}=4977$ MeV, $m_{2}=1628$ MeV, $m_{3}=220$ MeV)} & \multicolumn{6}{c}{Type III ($m_{1}=m_{2}=4977$ MeV, $m_{3}=220$ MeV)}\\
$(0,0)1S(\frac{1}{2}^{+})_{0}$ & 0.370  & 0.479 & - & - &- & $(0,0)1S(\frac{1}{2}^{+})_{1}$ &  0.297  & 0.470  & - & -  &-\\
           -             &    -         & -   & -& -  &- & $(0,0)1S(\frac{3}{2}^{+})_{1}$ &  0.299 & 0.483 & - & - &- \\
$(0,1)1P(\frac{1}{2}^{-})_{1}$ & 0.391  & $\mathbf{0.654}$ & $m_{q}\langle r_{\lambda}^{2}\rangle$ & $\sim412$ & 356 & $(0,1)1P(\frac{1}{2}^{-})_{1}$ & 0.313 & $\mathbf{0.666}$ & $m_{q}\langle r_{\lambda}^{2}\rangle$ & $\sim398$ &361\\
$(0,1)1P(\frac{3}{2}^{-})_{1}$ & 0.395  & $\mathbf{0.696}$ & $m_{q}\langle r_{\lambda}^{2}\rangle$ & $\sim364$ &356& $(0,1)1P(\frac{3}{2}^{-})_{1}$ & 0.314 & $\mathbf{0.688}$ & $m_{q}\langle r_{\lambda}^{2}\rangle$ & $\sim373$ &361 \\
$(1,0)1P(\frac{1}{2}^{-})_{1}$ & $\mathbf{0.562}$ & 0.503 & $\frac{1}{4}m_{Q}\langle r_{\rho}^{2}\rangle$ & $\sim99$ & 273& $(1,0)1P(\frac{1}{2}^{-})_{1}$ & $\mathbf{0.456}$ & 0.496 & $\frac{1}{2}m_{Q}\langle r_{\rho}^{2}\rangle$ & $\sim75$ & 237\\
$(1,0)1P(\frac{3}{2}^{-})_{1}$ & $\mathbf{0.566}$ & 0.522 & $\frac{1}{4}m_{Q}\langle r_{\rho}^{2}\rangle$ & $\sim97$ & 273 & $(1,0)1P(\frac{3}{2}^{-})_{1}$ & $\mathbf{0.459}$ & 0.508 & $\frac{1}{2}m_{Q}\langle r_{\rho}^{2}\rangle$ & $\sim74$ & 237\\
\end{tabular}
\end{ruledtabular}
\end{table*}

\begin{table*}[htbp]
\begin{ruledtabular}\caption{Contrast of the calculated results~\cite{f11,f11p} and data for the well established singly heavy baryons~\cite{f14}.}
\label{t3}
\begin{tabular}{c c c c c | c c c c c}
Baryon & $M_{exp.}$ & $J^{P}$ & $M_{cal.}$ & $nL(J^{P})$& Baryon & $M_{exp.}$ & $J^{P}$ & $M_{cal.}$ & $nL(J^{P})$  \\ \hline
$\Lambda_{c}^{+}$ & 2286.46$\pm$0.14  & $\frac{1}{2}^{+}$ & 2288 & $1S(\frac{1}{2}^{+})$          &$\Xi_{c}(2970)^{+}$ & 2964.3$\pm$1.5  & $\frac{1}{2}^{+}$ & 2949 & $2S(\frac{1}{2}^{+})(\bar{\mathbf{3}}_{F})$ \\
$\Lambda_{c}(2765)^{+}$~\cite{f14p} & 2766.6$^{+3.6}_{-7.1}$ & $\frac{1}{2}^{+}$ & 2764 & $2S(\frac{1}{2}^{+})$ & $\Xi_{c}(2970)^{0}$ & 2967.1$\pm$1.7 & $\frac{1}{2}^{+}$ & 2949 & $2S(\frac{1}{2}^{+})(\bar{\mathbf{3}}_{F})$ \\
$\Lambda_{c}(2595)^{+}$ & 2592.25$\pm$0.28  & $\frac{1}{2}^{-}$ & 2596 & $1P(\frac{1}{2}^{-})$   & $\Omega_{c}^{0}$ & 2695.2$\pm$1.7  & $\frac{1}{2}^{+}$ & 2699 & $1S(\frac{1}{2}^{+})$ \\
$\Lambda_{c}(2625)^{+}$ & 2628.11$\pm$0.19  & $\frac{3}{2}^{-}$ & 2631 & $1P(\frac{3}{2}^{-})$   & $\Omega_{c}(2770)^{0}$ & 2765.9$\pm$0.28  & $\frac{3}{2}^{+}$ & 2762 & $1S(\frac{3}{2}^{+})$ \\
$\Lambda_{c}(2860)^{+}$ & 2856.1$^{+2.3}_{-6.0}$  & $\frac{3}{2}^{+}$ & 2875 & $1D(\frac{3}{2}^{+})$   & $\Lambda_{b}^{0}$ & 5619.60$\pm$0.17  & $\frac{1}{2}^{+}$ & 5622 & $1S(\frac{1}{2}^{+})$ \\
$\Lambda_{c}(2880)^{+}$ & 2881.63$\pm$0.24  & $\frac{5}{2}^{+}$ & 2891 & $1D(\frac{5}{2}^{+})$  & $\Lambda_{b}(5912)^{0}$ & 5912.19$\pm$0.17  & $\frac{1}{2}^{-}$ & 5898 & $1P(\frac{1}{2}^{-})$ \\
$\Sigma_{c}(2455)^{++}$ & 2453.97$\pm$0.14  & $\frac{1}{2}^{+}$ & 2457 & $1S(\frac{1}{2}^{+})$     & $\Lambda_{b}(5920)^{0}$ & 5920.09$\pm$0.17  & $\frac{3}{2}^{-}$ & 5913 & $1P(\frac{3}{2}^{-})$ \\
$\Sigma_{c}(2455)^{+}$ & 2452.65$^{+0.22}_{-0.16}$  & $\frac{1}{2}^{+}$ & 2457 & $1S(\frac{1}{2}^{+})$ & $\Lambda_{b}(6070)^{+}$ & 6072.3$\pm$2.9  & $\frac{1}{2}^{+}$ & 6041 & $2S(\frac{1}{2}^{+})$ \\
$\Sigma_{c}(2455)^{0}$ & 2453.75$\pm$0.14  & $\frac{1}{2}^{+}$ & 2457 & $1S(\frac{1}{2}^{+})$      & $\Lambda_{b}(6146)^{0}$ & 6146.2$\pm$0.4  & $\frac{3}{2}^{+}$ & 6137 & $1D(\frac{3}{2}^{+})$ \\
$\Sigma_{c}(2520)^{++}$ & 2518.41$^{+0.22}_{-0.18}$  & $\frac{3}{2}^{+}$ & 2532 & $1S(\frac{3}{2}^{+})$ & $\Lambda_{b}(6152)^{0}$ & 6152.5$\pm$0.4  & $\frac{5}{2}^{+}$ & 6145 & $1D(\frac{5}{2}^{+})$ \\
$\Sigma_{c}(2520)^{+}$ & 2517.4$^{+0.7}_{-0.5}$  & $\frac{3}{2}^{+}$ & 2532 & $1S(\frac{3}{2}^{+})$ & $\Sigma_{b}^{+}$ & 5810.56$\pm$0.25  & $\frac{1}{2}^{+}$ & 5820 & $1S(\frac{1}{2}^{+})$ \\
$\Sigma_{c}(2520)^{0}$ & 2518.48$\pm$0.20  & $\frac{3}{2}^{+}$ & 2532 & $1S(\frac{3}{2}^{+})$     & $\Sigma_{b}^{-}$ & 5815.64$\pm$0.27  & $\frac{1}{2}^{+}$ & 5820 & $1S(\frac{1}{2}^{+})$ \\
$\Xi_{c}^{+}$ & 2467.71$\pm$0.23  & $\frac{1}{2}^{+}$ & 2479 & $1S(\frac{1}{2}^{+})(\bar{\mathbf{3}}_{F})$   & $\Sigma_{b}^{*+}$ & 5830.32$\pm$0.27  & $\frac{3}{2}^{+}$ & 5849 & $1S(\frac{3}{2}^{+})$ \\
$\Xi_{c}^{0}$ & 2470.44$\pm$0.28  & $\frac{1}{2}^{+}$ & 2479 & $1S(\frac{1}{2}^{+})(\bar{\mathbf{3}}_{F})$   & $\Sigma_{b}^{*-}$ & 5834.74$\pm$0.30  & $\frac{3}{2}^{+}$ & 5849 & $1S(\frac{3}{2}^{+})$ \\
$\Xi_{c}^{'+}$ & 2578.2$\pm$0.5  & $\frac{1}{2}^{+}$ & 2590 & $1S(\frac{1}{2}^{+})(\mathbf{6}_{F})$         & $\Xi_{b}^{-}$ & 5797.0$\pm$0.6  & $\frac{1}{2}^{+}$ & 5806 & $1S(\frac{1}{2}^{+})(\bar{\mathbf{3}}_{F})$ \\
$\Xi_{c}^{'0}$ & 2578.7$\pm$0.5  & $\frac{1}{2}^{+}$ & 2590 & $1S(\frac{1}{2}^{+})(\mathbf{6}_{F})$         & $\Xi_{b}^{0}$ & 5791.9$\pm$0.5  & $\frac{1}{2}^{+}$ & 5806 & $1S(\frac{1}{2}^{+})(\bar{\mathbf{3}}_{F})$ \\
$\Xi_{c}(2645)^{+}$ & 2645.1$\pm$0.3  & $\frac{3}{2}^{+}$ & 2658 & $1S(\frac{3}{2}^{+})(\mathbf{6}_{F})$    & $\Xi_{b}(5935)^{-}$ & 5935.02$\pm$0.05  & $\frac{1}{2}^{+}$ & 5943 & $1S(\frac{1}{2}^{+})(\mathbf{6}_{F})$ \\
$\Xi_{c}(2645)^{0}$ & 2646.16$\pm$0.25  & $\frac{3}{2}^{+}$ & 2658 & $1S(\frac{3}{2}^{+})(\mathbf{6}_{F})$   & $\Xi_{b}(5945)^{0}$ & 5952.3$\pm$0.6  & $\frac{3}{2}^{+}$ & 5971 & $1S(\frac{3}{2}^{+})(\mathbf{6}_{F})$ \\
$\Xi_{c}(2790)^{+}$ & 2791.9$\pm$0.5  & $\frac{1}{2}^{-}$ & 2789 & $1P(\frac{1}{2}^{-})(\bar{\mathbf{3}}_{F})$ & $\Xi_{b}(5955)^{-}$ & 5955.33$\pm$0.13  & $\frac{3}{2}^{+}$ & 5971 & $1S(\frac{3}{2}^{+})(\mathbf{6}_{F})$ \\
$\Xi_{c}(2790)^{0}$ & 2793.9$\pm$0.5  & $\frac{1}{2}^{-}$ & 2789 & $1P(\frac{1}{2}^{-})(\bar{\mathbf{3}}_{F})$ & $\Xi_{b}(6100)^{-}$ & 6100.3$\pm$0.6  & $\frac{3}{2}^{-}$ & 6097 & $1P(\frac{3}{2}^{-})(\bar{\mathbf{3}}_{F})$ \\
$\Xi_{c}(2815)^{+}$ & 2816.51$\pm$0.25  & $\frac{3}{2}^{-}$ & 2819 & $1P(\frac{3}{2}^{-})(\bar{\mathbf{3}}_{F})$ & $\Omega_{b}^{-}$ & 6045.2$\pm$1.2  & $\frac{1}{2}^{+}$ & 6043 & $1S(\frac{1}{2}^{+})$ \\
$\Xi_{c}(2815)^{0}$ & 2819.79$\pm$0.30  & $\frac{3}{2}^{-}$ & 2819 & $1P(\frac{3}{2}^{-})(\bar{\mathbf{3}}_{F})$ & $\Lambda_{c}(2940)^{+}$ & 2939.6$^{+1.3}_{-1.5}$   & $\frac{3}{2}^{-}$ & (?) & (?)  \\
\end{tabular}
\end{ruledtabular}
\end{table*}

\begin{table*}[htbp]
\begin{ruledtabular}\caption{The observed baryons with unconfirmed $J^{P}$ and the corresponding calculated results~\cite{f11,f11p}.}
\label{t4}
\begin{tabular}{c c c c| c c c c }
Baryon & $M_{exp.}$ &  $M_{cal.}$ & $nL(J^{P})$& Baryon & $M_{exp.}$ & $M_{cal.}$ & $nL(J^{P})$   \\ \hline
$\Sigma_{c}(2800)^{++}$~\cite{f14} & 2801$^{+4}_{-6}$  &  & $1P(\mathbf{6}_{F})$   & $\Sigma_{b}(6097)^{+}$~\cite{f14} & 6095.8$\pm$1.7   & &$1P(\mathbf{6}_{F})$ \\
$\Sigma_{c}(2800)^{+}$~\cite{f14} & 2792$^{+14}_{-5}$  &  & $1P(\mathbf{6}_{F})$ & $\Sigma_{b}(6097)^{-}$~\cite{f14} & 6098.0$\pm$1.8    & &$1P(\mathbf{6}_{F})$ \\
$\Sigma_{c}(2800)^{0}$~\cite{f14} & 2806$^{+5}_{-7}$   &  & $1P(\mathbf{6}_{F})$  & $\Xi_{b}(6227)^{-}$~\cite{f14} & 6227.9$\pm$0.9   & & $1P(\mathbf{6}_{F})$ \\
                      &                &    &        & $\Xi_{b}(6227)^{0}$~\cite{f14} & 6226.8$\pm$1.6   & & $1P(\mathbf{6}_{F})$ \\\hline
$\Xi_{c}(3055)^{+}$~\cite{f14} & 3055.9$\pm$0.4 &  3063 & $1D(\frac{3}{2}^{+})(\bar{\mathbf{3}}_{F})$ & $\Xi_{b}(6327)^{0}$~\cite{f14} & 6327.28$\pm$0.35 & 6320 & $1D(\frac{3}{2}^{+})(\bar{\mathbf{3}}_{F})$   \\
$\Xi_{c}(3080)^{+}$~\cite{f14} & 3077.2$\pm$0.4 &  3076 & $1D(\frac{5}{2}^{+})(\bar{\mathbf{3}}_{F})$  & $\Xi_{b}(6333)^{0}$~\cite{f14} & 6332.69$\pm$0.28   & 6327 & $1D(\frac{5}{2}^{+})(\bar{\mathbf{3}}_{F})$  \\
$\Xi_{c}(3080)^{0}$~\cite{f14} & 3079.9$\pm$1.4 &  3076 & $1D(\frac{5}{2}^{+})(\bar{\mathbf{3}}_{F})$ &    &         &    &    \\\hline
$\Omega_{c}(3000)$~\cite{f14} & 3000.41$\pm$0.22 &  & $1P(\mathbf{6}_{F})$  & $\Omega_{b}(6316)^{-}$~\cite{f14} & 6315.6$\pm$0.6   & & $1P(\mathbf{6}_{F})$  \\
$\Omega_{c}(3050)^{+}$~\cite{f14} & 3050.19$\pm$0.13 &  & $1P(\mathbf{6}_{F})$& $\Omega_{b}(6330)^{-}$~\cite{f14} & 6330.3$\pm$0.6   & & $1P(\mathbf{6}_{F})$  \\
$\Omega_{c}(3065)^{0}$~\cite{f14} & 3065.54$\pm$0.26  & & $1P(\mathbf{6}_{F})$ & $\Omega_{b}(6340)^{-}$~\cite{f14} & 6339.7$\pm$0.6   & & $1P(\mathbf{6}_{F})$\\
$\Omega_{c}(3090)^{0}$~\cite{f14} & 3090.1$\pm$0.5   &  & $1P(\mathbf{6}_{F})$ & $\Omega_{b}(6350)^{-}$~\cite{f14} & 6349.8$\pm$0.6    & & $1P(\mathbf{6}_{F})$\\\hline
$\Xi_{c}(2923)^{0}$~\cite{f16} & 2923.04$\pm$0.59   & & $1P(\mathbf{6}_{F})$ & $\Omega_{c}(3185)^{0}$~\cite{f16p3} & 3185.1$\pm$1.7   & 3197 & $2S(\frac{3}{2}^{+})(\mathbf{6}_{F})$ \\
$\Xi_{c}(2939)^{0}$~\cite{f16} & 2938.55$\pm$0.52  & & $1P(\mathbf{6}_{F})$   & $\Omega_{c}(3327)^{0}$~\cite{f16p3} & 3327.1$\pm$1.2   & & $1D(\mathbf{6}_{F})$\\
$\Xi_{c}(2964)^{0}$~\cite{f16} & 2964.88$\pm$0.54   & & $1P(\mathbf{6}_{F})$   & $\Xi_{b}(6095)^{0}$~\cite{f16p4} & 6095.36$\pm$0.68   & 6097 & $1P(\frac{3}{2}^{-})(\bar{\mathbf{3}}_{F})$ \\
$\Xi_{c}(2930)^{+}$~\cite{f16p1} & 2942.3$\pm$5.9    & & $1P(\mathbf{6}_{F})$ & $\Xi_{b}(6087)^{0}$~\cite{f16p4} & 6087.24$\pm$0.76   & 6084 & $1P(\frac{1}{2}^{-})(\bar{\mathbf{3}}_{F})$  \\\hline
$\Xi_{c}(3123)^{+}$~\cite{f16p2} & 3122.9$\pm$1.3   & (?) & (?)  & $\Omega_{c}(3120)^{+}$~\cite{f14} & 3119.1$\pm$1.0 & (?) & (?)  \\
\end{tabular}
\end{ruledtabular}
\end{table*}

\end{document}